\documentclass{article}
\usepackage{spconf,amsmath,graphicx}
\usepackage{url}

\title{Multi-task Learning for Voice Trigger Detection}
\name{Siddharth Sigtia, Pascal Clark, Rob Haynes, Hywel Richards, John Bridle}
\address{Apple}
\begin{document}
%
\maketitle
\begin{abstract}


We describe the design of a voice trigger detection system for smart speakers. In this study, we address two major challenges. The first is that the detectors are deployed in complex acoustic environments with external noise and loud playback by the device itself. Secondly, collecting training examples for a specific keyword or trigger phrase is challenging resulting in a scarcity of trigger phrase specific training data. We describe a two-stage cascaded architecture where a low-power detector is always running and listening for the trigger phrase. If a detection is made at this stage, the candidate audio segment is re-scored by larger, more complex models to verify that the segment contains the trigger phrase. In this study, we focus our attention on the architecture and design of these second-pass detectors. We start by training a general acoustic model that produces phonetic transcriptions given a large labelled training dataset. Next, we collect a much smaller dataset of examples that are challenging for the baseline system. We then use multi-task learning to train a model to simultaneously produce accurate phonetic transcriptions on the larger dataset \emph{and} discriminate between true and easily confusable examples using the smaller dataset. Our results demonstrate that the proposed model reduces errors by half compared to the baseline in a range of challenging test conditions \emph{without} requiring extra parameters.


\end{abstract}
\begin{keywords}
Keyword Spotting, Acoustic Modelling, Multi-task Learning, Neural Networks
\end{keywords}
\section{Introduction}
\label{sec:intro}

There has been a proliferation in the use of speech-based services with \emph{voice-first} user interfaces like Siri. For smart speakers, speech is the primary means of interaction. Typically a specific keyword or \emph{trigger phrase} is used to initiate an interaction with the device. For instance in English, the phrase \emph{Hey Siri} is used for all Apple devices. Given that the trigger phrase acts as an on-switch for user interactions, the voice trigger detection algorithm must be very accurate. Smart speakers pose a particularly challenging problem for the design of voice trigger detection algorithms since they are deployed in acoustically challenging environments. The user can be far away from the device which results in weak and reverberated speech. The detectors must work accurately in the presence of noise from sources like TVs, radios and household appliances. Furthermore, the detectors must work during loud playback from the device itself, which results in the amplitude of the music being significantly greater than the user's speech. Although modern echo cancellation algorithms are able to successfully remove most of the playback signal, non-linearities and non-stationarities in the audio rendering process result in residual echo which still poses a challenge for both trigger detection and speech recognition \cite{MLBlogFrontEnd}. Another significant challenge is that unlike automatic speech recognition (ASR) systems, collecting training examples for a specific keyword or phrase in a variety of conditions is a difficult problem (c.f. Section 4).  


\begin{figure}[t]
\begin{minipage}[]{1.0\linewidth}
  \centering
  \centerline{\includegraphics[width=0.8\textwidth]{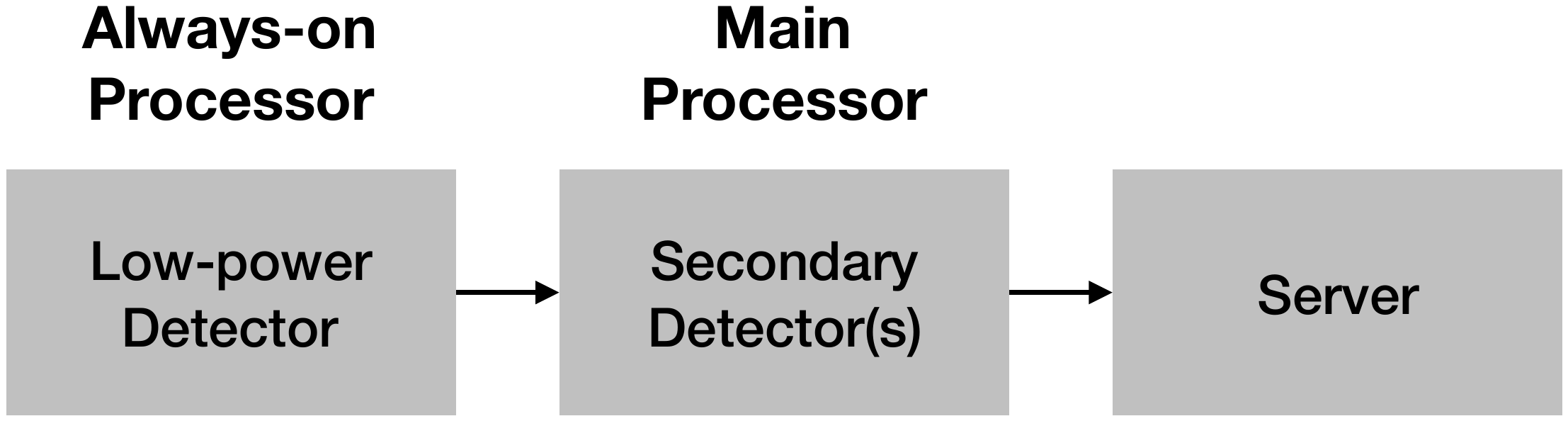}}
  \caption{Overview of the two-stage architecture \cite{MLBlogHS,sigtia2018}.}
  \label{block_diagram}
\end{minipage}
\end{figure}

In the literature, the problem of detecting a speech trigger phrase is interchangeably referred to as voice trigger detection \cite{sigtia2018}, keyword spotting \cite{chen2014small}, wake-up word detection \cite{kumatani2017direct} or hotword detection \cite{huang2019multi}. In the rest of this paper, we refer to this problem as voice trigger detection. Recent approaches to trigger detection use various neural network architectures like DNNs \cite{sigtia2018,chen2014small,choi2019temporal}, RNNs \cite{fernandez2007application,he2017streaming,yamamoto2019small} and CNNs \cite{sainath2015convolutional,arik2017convolutional,kao2019sub}. Another way to characterise voice trigger detection algorithms is whether they use a single pass approach \cite{chen2014small} or a two-stage cascaded architecture \cite{MLBlogHS,sigtia2018,gruenstein2017cascade,wu2018monophone}. In the multi-stage approach (Figure \ref{block_diagram}), the first stage comprises a low-power DNN-HMM system that is always on \cite{sigtia2018}. If a detection is made at this stage, the acoustic segment is passed on to larger, more complex models which are used to re-score the segment by calculating the probability of the trigger phrase given the acoustic evidence. In this design, it is the second stage that determines the final accuracy of the system and the models used in this stage are the subject of this paper.

In this study we address the major challenges outlined above i.e.\ trigger detection in the presence of external noise, reverberated speech and echo residuals and the issue of scarcity of labelled training data for a given phrase. Our \textbf{main contribution} is to propose a multi-task learning strategy where a single model is trained to optimise 2 objectives simultaneously. The first objective is to assign the highest score to the correct sequence of phonetic labels given a speech recording. This objective is optimised on a large labelled training dataset which is also used for training the main speech recogniser and is therefore easy to obtain. The second objective is to discriminate between utterances that contain the trigger phrase and those that are phonetically similar and easily confusable. This dataset is significantly smaller compared to the ASR training set. Our results demonstrate that this strategy yields significant gains on two challenging test sets, reducing the number of errors by half in some conditions, without requiring extra parameters compared to the baseline system. 

\vspace{-6mm}
\section{Baseline}
\label{sec:baseline}

\vspace{-2mm}
The baseline model architecture comprises an acoustic model (AM) with four bidirectional LSTM layers with 256 units each, followed by an output affine transformation + softmax layer over context independent (CI) phonemes, word and sentence boundaries, resulting in 53 output symbols (Figure 2). The model contains roughly five million parameters. The AM is trained by minimising the Connectionist Temporal Classification (CTC) loss function \cite{graves2006connectionist}. The inputs to the model are 40-dimensional Mel-filterbank features computed at 100 frames-per-second (FPS) which are then stacked to form symmetric input windows of 7 frames. The sequence of windows is then sub-sampled by a factor of 3. This choice of model architecture offers several advantages. Firstly, the fact that the second-pass model is used for re-scoring and not in a continuous streaming setting allows us to use bidirectional LSTM layers. Secondly, using context-independent phones as targets allows us to share training data with the main ASR. This is particularly important since in many cases it is not possible to obtain a large number of training utterances with the trigger phrase, for example when developing a trigger detector for a new language. Furthermore, having CI phones as targets results in a flexible model that can be used for detecting any keyword. Next, AMs trained with CTC are operated at 33 frames-per-second (FPS) as opposed to most DNN-HMM systems that are operated at 100 FPS. This reduces the amount of computation performed on-device by a factor of 3. Finally for inference, given an audio segment $\mathbf x$ from the first pass, we are interested in calculating the probability of the phone sequence in the trigger phrase, $P(\text{TriggerPhrasePhoneSeq}|\mathbf x)$. This computation can be compactly expressed as a left-to-right HMM \cite{hannun2017sequence}. Consequently, the forward probabilities can be efficiently computed using dynamic programming.

\section{Multi-task Learning}

The model described above suffers from an obvious shortcoming: the training objective differs from the final objective that we are interested in. The LSTM AM is trained to output the correct sequence of CI phonemes given an input speech utterance. While at inference, we calculate the probability of the phone sequence in the trigger phrase given the acoustic evidence, $P(\text{TriggerPhrasePhoneSeq}|\mathbf x)$. However the question we really want to answer is, ``given an audio segment from the first pass, does it contain the trigger phrase or not?'' The LSTM AM proposed above expends a lot of capacity towards modelling phonemes we do not care about. And unsurprisingly, examples that are phonetically similar to the trigger phrase are assigned high scores which results in false detections. 

Ideally, we would like the second-pass model to be a binary classifier which determines the presence or absence of the trigger phrase. We could then collect a large number of examples of each class and optimise the correct objective function during training. However the issue with this design is that collecting a large number of training examples that result in false detections by the baseline system is a difficult problem (c.f. Section 4). Furthermore, the second pass models have millions of parameters, so they can easily overfit a small training set resulting in poor generalisation. Therefore, we are faced with the choice between a more general phonetic AM that can be trained on a large, readily available dataset but is optimised for the wrong criterion \emph{or} a trigger phrase specific detector that is trained on the correct criterion but with a significantly smaller training dataset.

\begin{figure}[t]
\begin{minipage}[]{1.0\linewidth}
  \centering
  \centerline{\includegraphics[width=0.8\textwidth]{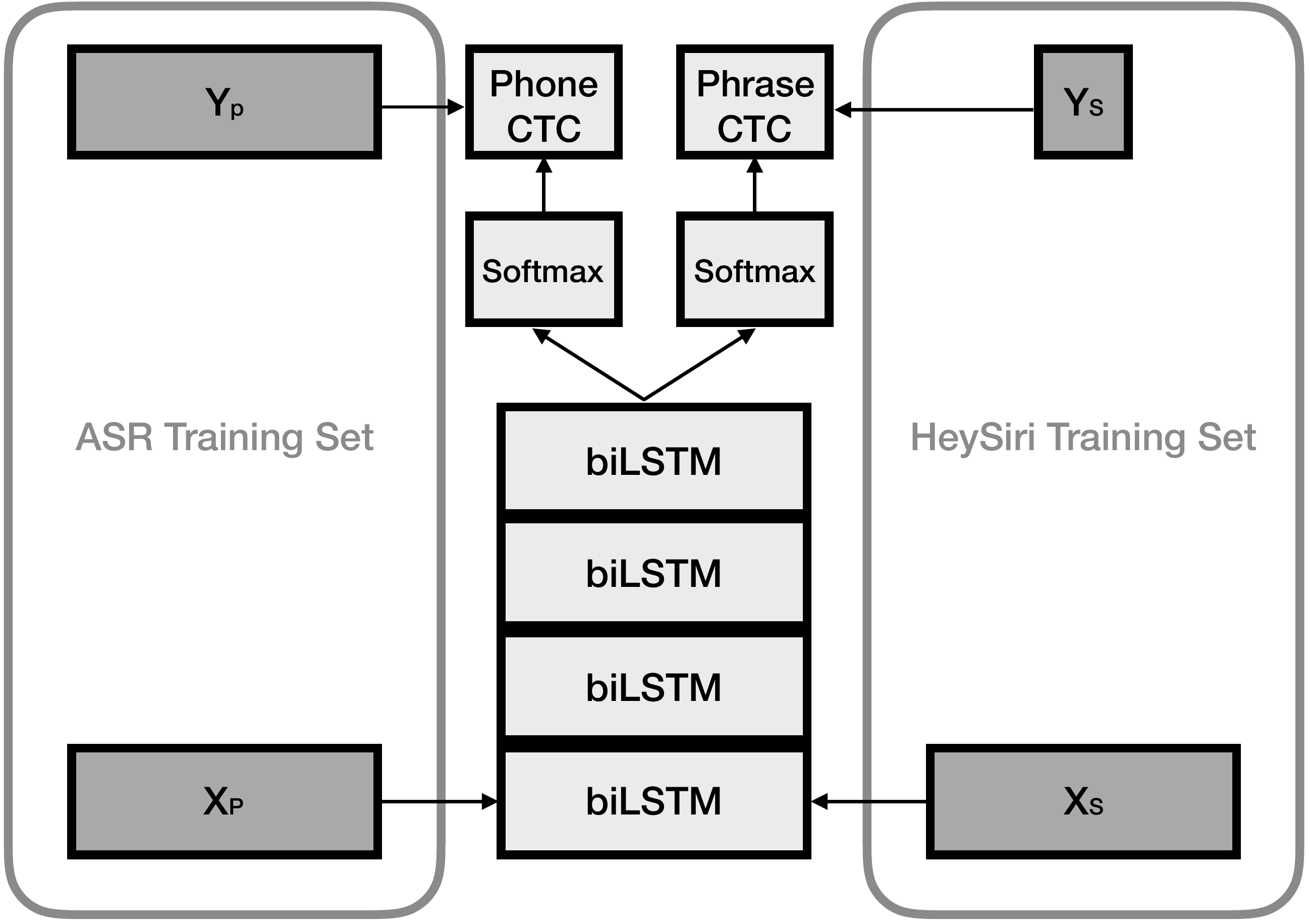}}
  \caption{Multi-task learning setup. The network comprises 4x shared bi-directional LSTM layers with tied weights. The CTC loss function is jointly optimised for both tasks. Note that the training sets are disjoint.}
  \label{multitask}
\end{minipage}
\end{figure}

One solution to this problem is to use multi-task learning (MTL) \cite{caruana1997multitask,huang2015rapid}. Intuitively, the main idea is that for tasks that are related, a single network should be able to learn \emph{shared} representations that are able to exploit the commonalities in the given tasks to yield better/more useful representations than training networks for each task alone. Note that predicting the sequence of phonetic labels in an utterance and deciding whether an utterance contains a specific trigger phrase or not, are \emph{related} tasks. The representations learnt by the network to solve one task should be useful to solve the other task. Therefore rather than training two separate networks for each task, we train a single network with a stack of shared/tied biLSTM layers with two seperate output layers (one for each task) and train the network jointly on both sets of training data (Figure \ref{multitask}). We hypothesise that the joint network is able to learn useful features from both tasks: a) the network can be trained to predict phone labels on a large labelled dataset of general speech which covers a wide distribution of complex acoustic conditions, b) the same network can also learn to discriminate between examples of true triggers and confusable examples on a relatively smaller dataset. An alternative view of this process is that the phonetic transcription task with a significantly larger training set acts as a regulariser for the trigger phrase discrimination task with a much smaller dataset. 


The objective function for the phrase specific/discriminative output layer is defined as follows: the softmax output layer contains two output units, one for the trigger phrase and the other one for the blank symbol used by the CTC loss function \cite{fernandez2007application,graves2006connectionist}. When the input $\mathbf x$ contains the trigger phrase, we minimise the CTC loss for the target label sequence $\mathbf y = \{\epsilon,\text{TriggerPhrase},\epsilon\}$. When the input segment does not contain the trigger phrase, the target label sequence is $\mathbf y = \{\epsilon\}$, which is equivalent to minimising the cross-entropy loss $C = -\sum_{t} \log \hat y_{\epsilon t}$, where $\hat y_{\epsilon t}$ is the network output for the blank symbol at time-step $t$. At inference time given an input segment $\mathbf x$, the detection score is simply $P(\mathbf y|\mathbf x)$, where $\mathbf y = \{\epsilon,\text{TriggerPhrase},\epsilon\}$.
The  MTL objective function that is minimised is a linear combination of both objectives:
\begin{equation}
	C_{MTL}{\big(\theta_{Tied},\theta_{P},\theta_{D} \big)} = C_P{\big(\theta_{Tied},\theta_P \big)} + C_D{\big(\theta_{Tied},\theta_D \big)},
\end{equation}
where $C_P$ is the phonetic CTC objective function, $C_D$ is the discriminative/phrase specific CTC objective function, $\theta_{Tied}$ are the tied weights of the biLSTM layers, $\theta_P$ are the parameters of the output (fully connected + softmax) layer for the phonetic model and $\theta_D$ are the parameters of the discriminative/keyword specific model (Figure 2). 

\section{Training Data}

\subsection{Baseline Phonetic CTC Model}

The baseline phonetic CTC model is trained using a large dataset of manually transcribed utterances that are sampled from intended invocations of the voice assistant. We start with a dataset of roughly 1 million transcribed utterances, which contain over 1500 hours of audio. These examples are primarily recorded on mobile phones and tend to not include reverberant speech, users at a distance, and echo residuals, resulting in a mismatch between the training data and their intended use for smart speakers. To reduce this mismatch, we employ the following  data augmentation strategies. Firstly, we convolve the original dataset with room impulse responses (RIRs) to simulate reverberant audio. We use a set of 3000 RIRs internally collected in a wide variety of different rooms in various houses. Each utterance is convolved with a randomly selected RIR from this list. Next, we add echo residuals to the reverberated examples to simulate conditions where the device is playing audio. We collect 400,000 examples of echo residuals with the device playing music, podcasts and text-to-speeech (TTS) at various volumes. Each reverberated utterance is mixed with a randomly selected example from this list. The final dataset comprises the original \emph{clean} data, the reverberated data and the reverberated data with echo residuals, which results in approximately 5000 hours of labelled training data. 

\vspace{-3mm}
\subsection{Multi-task Training Data}

\begin{figure*}[htb]
\centering
\begin{minipage}[htb]{1.0\linewidth}
  \centering
  \centerline{\scalebox{1.}{\includegraphics[width=0.85\textwidth,height=6cm]{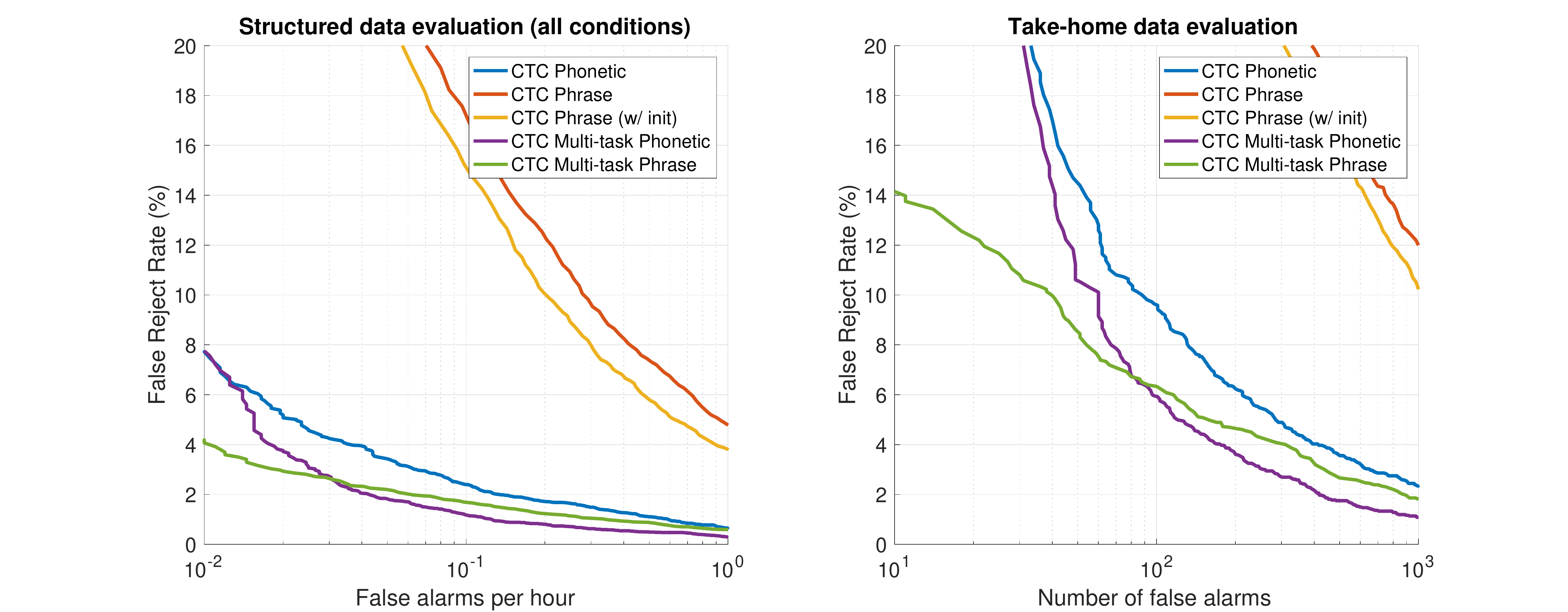}}}
  \caption{DET curves for structured evaluation set (left) and the take-home evaluation set (right).}
  \label{results}
\end{minipage}
\end{figure*}

As mentioned before, collecting difficult examples that falsely trigger the baseline system is challenging. We collect a dataset of 40,000 utterances that result in false triggers and 140,000 examples that contain the trigger phrase. We then extract the audio segments corresponding to the trigger phrase as determined by the first-pass DNN-HMM system. The total dataset comprises 90 hours of audio with binary labels (trigger phrase or not). Note that this dataset is two orders of magnitude smaller than the baseline training dataset. The multi-task training dataset is formed by concatenating the ASR dataset described above and the much smaller phrase specific dataset described here. 

\vspace{-3mm}
\section{Model Training}

Both the baseline and the multi-task models are trained using exactly the same optimiser and hyper-parameter settings. We use large-batch distributed training with synchronous gradient updates \cite{dean2012large}. We use a mini-batch of 128 utterances to compute the gradient per GPU and we use 32 GPUs in parallel. We use an initial learning rate of 0.0032 and the Adam optimiser to update the weights and hyper-parameters. In order to avoid gradient explosion at the start of training, we clip the norm of the gradient to a value of 5. 

\section{Evaluation}

In this section we present results on two internally-collected voice trigger test sets. Both were recorded on-device from live interactive sessions in realistic environments and conditions. The first test set is a structured data collection with controlled acoustic conditions. Recordings were made on-device while each subject made a series of prompted voice commands. There were 100 subjects, approximately balanced between male and female adults. Distances from the device were controlled, ranging from 8 to 15 feet away. There are over 13K utterances overall, evenly divided between four acoustic conditions: (a) quiet room, (b) external noise from a TV or kitchen appliance in the room, (c) music playback from the recording device at medium volume, and (d) music playback from the recording device at loud volume.  Note that condition (d) is quite challenging due to considerable levels of residual noise that the voice-trigger model must contend with in order to detect the trigger-phrase. These examples are used to measure the proportion of false rejections (FRs). In addition to these recordings, this test set also consists of almost 2,000 hours of continuous audio recordings from TV, radio, and podcasts. This allows the measurement of the false-alarm (FA) rate in terms of FA’s per hour of active external audio.

The second test set is an unstructured data collection at home by our employees, designed to be more representative of realistic, spontaneous usage of the smart speaker. For this data collection, each of the 42 participants took home a device and used it daily for two weeks. Extra audio logging on the device, and personal review by the user, allowed the collection of audio well below the usual on-device trigger threshold. Continuous audio recording was not possible, so instead there was a low-threshold first-pass detected that recorded audio segments acoustically similar to the trigger phrase. With this data, it is possible to measure nearly unbiased false-reject and false-alarm rates for realistic in-home scenarios similar to customer usage. It is not possible to use customer data directly, where the false-reject rate is unmeasurable due to the fact that the only audio sent to the server are utterances which have already triggered the device. 


The results are shown in Figure 3. We use detection-error trade-off (DET) curves to compare the accuracy between models. Each curve displays the FA rate and the proportion of FRs associated with sweeping the trigger threshold for a particular model. In practice, we compare the shapes of the DET curves for different models in the vicinity of viable operating points. We compare five models: the baseline phonetic CTC model trained on the ASR dataset (blue), the baseline phrase specific model trained on the much smaller training set with randomly initialised weights (red), the same phrase specific model but with weights initialised with the learned weights from the baseline phonetic CTC model (yellow), the phonetic (purple) and phrase specific (green) branches of the proposed MTL model. Note that the phrase specific model with weight initialisation from the baseline phonetic model (yellow) is effectively trained using both datasets. In both test sets, the MTL phonetic (purple) and phrase-specific (green) models outperform the baseline phonetic CTC (blue), reducing the FR rate by almost half at many points along the curve. Note that the two MTL models share all their parameters and differ only in the output layer. The non-MTL phrase specific models (red and yellow) yield significantly worse accuracies in comparison, which is unsurprising given that the training dataset is two orders of magnitude smaller compared to the phonetic baseline (blue). Comparing the structured data evaluation (left) and the take-home data evaluation (right), it is also striking how the error rates are generally much higher for the latter. Despite the presence of very loud device playback in the structured evaluation dataset, the take-home data is more challenging in terms of several factors. For instance, in-home data is more likely to be spontaneous and to contain non-stationary noise in the background. Also, there is a wider range of voice types from the in-home data collection, including children. Even on this challenging test set, the MTL models are able to reduce FR rates by half across a wide range of operating points. 


\section{Conclusions}

We presented an architecture for performing voice trigger detection on smart speakers in challenging acoustic environments. We introduced a multi-task learning strategy to adapt a general acoustic model for voice trigger detection. We trained the model to simultaneously produce phonetic transcriptions on a large ASR dataset and to discriminate between difficult examples on a much smaller trigger phrase specific training set. We evaluate the proposed model on two challenging test sets and find the proposed method is able to almost halve errors and does not require any extra model parameters. 


\bibliographystyle{IEEEbib}
\bibliography{refs}

\end{document}